\documentclass{elsart}

\newcommand {\be}{\begin{equation}}
\newcommand {\ee}{\end{equation}}
\newcommand{\bea}{\begin{eqnarray}}
\newcommand{\eea}{\end{eqnarray}}
\newcommand {\vett}[1] {\mathbf{#1}}

\begin{document}
\begin{frontmatter}
\title{Collective excitation frequencies of Bosons in a parabolic potential
with interparticle harmonic interactions}
\author[Pisa]{A. Minguzzi},
\author[Pisa]{M.P. Tosi} and
\author[Oxford]{N.H. March}
\address[Pisa]{Istituto Nazionale di Fisica della Materia and Classe
di Scienze,  Scuola Normale Superiore, Piazza dei Cavalieri 7, I-56126
Pisa, Italy} 
\address[Oxford]{Oxford University, Oxford, England and Department of
Physics, University of Antwerp (RUCA), Groenenborgerlaan, B-2020
Antwerpen, Belgium} 

\begin{abstract}
	The fact that the ground-state first-order density matrix for
	Bosons in a parabolic potential with interparticle harmonic
	interactions is known in exact form is here exploited to study
	collective excitations in the weak-coupling
	regime. Oscillations about the ground-state density are
	treated analytically by a linearized equation of motion which
	includes a kinetic energy contribution. We show that the
	dipole mode has the frequency of the bare trap, in accord with
	the Kohn theorem, and derive explicit expressions for the
	frequencies of the higher-multipole modes in terms of a
	frequency renormalized by the interactions. 


\end{abstract}

\end{frontmatter}

\section{Introduction}

	In a very recent investigation, Amoruso {\it et al.} [1] have given
	a comparative discussion of collective excitations in dilute
	Fermi and Bose gases subject to harmonic confinement at zero
	temperature. For this purpose these workers employed a
	linearized form of the equation of motion for the density
	profile $n(\vett R,t)$ in the Hartree approximation [2 - 4], which is 
\begin{eqnarray}
\frac{\partial^2 n(\vett R,t)}{\partial t^2}&=&\frac{1}{m}\nabla_{\vett
	R} \cdot \left\{n(\vett R,t)\nabla_{\vett R}\left
	[ V_{ext}(\vett R)+\int d^3 x \, v(\vett R- \vett x) n(\vett
	x,t)\right] \right\} \nonumber \\ &+&\frac 1 m
	\sum_{\alpha,\beta} \nabla_{\alpha}\nabla_{\beta}
	\Pi_{\alpha,\beta}(\vett R,t)
\end{eqnarray}	
where $V_{ext}(\vett R)=m \omega_{ho}^2 R^2/2$ is the confining
potential, $v(\vett R-\vett x)$ is the interparticle
interaction and $\Pi_{\alpha,\beta}(\vett R,t)$ is the kinetic
stress tensor. 

Whereas for a Bose-condensed cloud Amoruso {\it et al.} [1] chose the
interactions to be of contact form $v(\vett R-\vett x)=g \delta(\vett
R-\vett x)$ and adopted the Thomas-Fermi approximation (corresponding
to $\Pi_{\alpha,\beta}(\vett R,t)$ = 0), the focus here is the N-Boson model
discussed by earlier 
workers [5, 6]. In this model, all Boson properties are governed by
harmonic forces, both confinement (which is usual) and, in contrast to
the contact interactions in ref. [1], 
\be
v(\vett R-\vett x)=\pm \frac 1 2 \gamma^2 (\vett R-\vett x)^2 \;.
\ee
Notice that the positive (negative) sign in Eq.~(2) corresponds to
attractions (repulsions).

One merit of this model is that the equilibrium density profile
$n_0(\vett R)$ has the exact form
\be
n_0(\vett R)=N(\kappa_N/\pi)^{3/2}\exp(-\kappa_N R^2)
\ee
with
\be
\kappa_N=\frac{N \omega_{ho}\omega_N}{(N-1)\omega_{ho}+\omega_N }
\ee
and where
\be
\omega_N^2=\omega_{ho}^2\pm N \gamma^2
\ee
$N$ being the number of Bosons in the cloud. In the repulsive case one
is assuming that the interactions are not so strong as to overcome the
confinement. 
	
In this Letter we show that within this model it is possible to
determine in a self-consistent manner the equilibrium density profile
and the collective excitation spectrum of the full Eq.~(1) in the weak
coupling limit. The advantage of having analytic results is that they
can be used to study and test different ideas and numerical methods. 

\section{Equilibrium density profile at weak coupling}

We have already pointed out that Eq.~(1) involves the Hartree
approximation, corresponding to the neglect of quantal fluctuations in
treating the potential energy terms. We introduce a parallel treatment
of the kinetic stress tensor, by using a decoupling approximation on
the one-body density matrix in terms of the product of two condensate
wave functions. Namely, we write $\langle\psi^\dag(\vett x',t) \psi
(\vett x'',t)\rangle\simeq \Phi^*(\vett x',t)\Phi(\vett x'',t)$ and
correspondingly find  
\be
\sum_{\alpha,\beta} \nabla_{\alpha}\nabla_{\beta}
\Pi_{\alpha,\beta}(\vett R,t)=\frac{-\hbar^2}{2m}\nabla_{\vett R}
\cdot \left\{ n(\vett R,t) \nabla_{\vett
R}\left[\frac{1}{\sqrt{n(\vett R,t)}}\nabla^2\sqrt{n(\vett R,t)}
\right]\right\} \,.
\ee
In the same approximation the Boson density is given by $n(\vett
R,t)=|\Phi(\vett R,t)|^2$, {\it i.e.} we
neglect the depletion of the condensate due to the interactions. This
approximate scheme has been shown to be equivalent to adopting the
Gross-Pitaevskii equation for the condensate [4].

In this approximation the equilibrium density profile $n_0(\vett R)$
is the solution of the following equation,
\be
V_{ext}(\vett R)\pm \frac{\gamma^2}{2} \int d^3x \, (\vett R- \vett x)^2
n_0(\vett x)-\frac{\hbar^2}{2m \sqrt{n_0(\vett R)}} \nabla^2
\sqrt{n_0(\vett R)}=\mu\;,
\ee
$\mu$ being the chemical potential. We choose as an Ansatz for the
solution of Eq.~(7) a gaussian profile normalized to the total number
of particles: 
\be
n_0(\vett R)=N (\kappa/\pi)^{3/2} \exp(-\kappa R^2)\;.
\ee	
Upon substitution of Eq.~(8) in Eq.~(7) and equating to zero the
coefficients of the $R^0$ and $R^2$ terms we find
\be
\mu=\frac{3}{2} \left( \kappa \mp \frac{N \gamma^2}{\kappa}\right)
\ee
and
\be
\kappa^2=\omega_{ho}^2 \pm N \gamma^2  \;.
\ee
Of course, the width of the equilibrium profile is narrowed
(broadened) by attractive (repulsive) interactions.

We can now compare the approximate result in Eqs.~(8) and (10)
with the exact result in Eqs.~(3) - (5). It is seen at once
that the two results agree in the weak coupling limit ($\omega_{ho}^2
\gg N \gamma^2$),
where Eqs.~(4) and (5) yield $\kappa_N^2\simeq \omega_{ho}^2 \pm N
\gamma^2 $ (we are neglecting unity 
relative to N). This comparison emphasizes the limits of
validity of our approach, which involves the neglect of
quantum fluctuations and of the depletion of the condensate.

\section{ Collective excitations at weak coupling}

We proceed to evaluate the small-amplitude oscillations of the Bose
cloud around the approximate density profile given in Eq.~(8). We set
$n(\vett R,t)=n_0(\vett R)+n_1(\vett R,t)$ in the equation of motion
(1) with the approximate form of the 
kinetic stress tensor in Eq.~(6) and linearize it in the fluctuation
$n_1(\vett R,t)$. We emphasize that this procedure is treating the equilibrium
profile and the dynamic fluctuations of the profile in a consistent
manner. The linearized equation of motion is 
\bea
\frac{\partial^2 n_1(\vett R,t)}{\partial t^2}&=&\frac{1}{m}
\nabla_{\vett R} \cdot \left\{n_0(\vett R)\nabla_{\vett R} \left[\pm
\frac{\gamma^2}{2} \int d^3x\, (\vett R-\vett x)^2 n_1(\vett
x,t)\right]\right\}\nonumber\\ &-&\frac{\hbar^2}{4 m^2}\nabla_{\vett
R}\cdot \left\{ n_0(\vett R) \nabla_{\vett R} \left[\frac{1}{\sqrt{n_0(\vett 
R)}}\nabla^2 \frac{n_1(\vett
R,t)}{\sqrt{n_0(\vett R)}}\right.\right. \nonumber \\
&&-\left.\left.\frac{n_1(\vett R,t)}{(n_0(\vett R))^{3/2}} \nabla^2 
\sqrt{n_0(\vett R)}\right]\right\}\;.
\eea
Carrying out a Fourier transform with respect to the time variable and
using Eq.~(8), the equation obeyed by $n_1(\vett R,\omega)$ becomes
\bea
\omega^2 n_1(\vett R,\omega)&=&\mp 2 \gamma^2 \kappa n_0(\vett R) \vett R
\cdot \vett d (\omega)\nonumber \\ &+&\frac 1 4 \left\{ \nabla_{\vett
R}^2 A(\vett 
R,\omega) +\frac{2 \kappa}{R^2}\frac{\partial}{\partial R}\left[R^3
A(\vett R,\omega)\right]\right\}
\eea
where
\be
A(\vett R,\omega)=\nabla_{\vett R}^2 n_1(\vett R,\omega) + 2 \kappa R
\frac{\partial}{\partial R}n_1(\vett R,\omega)+ 6 \kappa n_1(\vett R,\omega)
\ee
and we have defined the dipole moment of the excitation as
\be
\vett d (\omega)=\int d^3x \, \vett x n_1(\vett x,\omega) \;.
\ee
Furthermore, we have used rescaled units such that $m=1$ and $\hbar=1$.

We now focus first of all on the equation of motion for a dipolar
density fluctuation, which is obtained from
Eq.~(12) by taking its dipole moment. Setting
\be
n_1(\vett R,\omega)\propto R Y_{1m}(\theta,\phi)\exp(-\kappa R^2)
\ee
in this case, it is easily seen that $A(\vett R,\omega)\propto
n_1(\vett R,\omega)$. The average of the Hartree term is
\be
\int d^3R \, R_{\alpha}R_{\beta}n_0(\vett R)=(N/2\kappa) \delta_{\alpha\beta}\,
\ee
while the average of the contribution from the kinetic stress tensor yields
\be
\frac 1 4 \int d^3R\, R_{\alpha} \left\{ \nabla_{\vett R}^2 A(\vett
R,\omega) +\frac{2 \kappa}{R^2}\frac{\partial}{\partial R}\left[R^3
A(\vett R,\omega) \right]\right\} = \kappa^2 d_{\alpha}(\omega)\;. 
\ee
The equation of motion of the dipole therefore takes the form
\be
\omega^2 d_{\alpha}(\omega)= (\kappa^2\mp N \gamma^2)d_{\alpha}(\omega)\;.
\ee

From Eqs.~(18) and (10) we see that the dipole mode oscillates at the
bare trap frequency, {\it i.e.} $\omega=\omega_{ho}$, in agreement with the Kohn theorem [7 -
9]. The theorem would be violated if we used the exact equilibrium
profile (3) - (5) in the approximate equation of motion (11).

We next turn to consider the dispersion relation for all the other
modes. What is shown below is that this dispersion behaviour for the
higher-multipole modes is given by multiples of the renormalized
harmonic oscillator frequency $\kappa$.

We begin by taking as an Ansatz for the density fluctuations about the
spherical equilibrium profile the form
\be
n_1(\vett R,\omega)=\Phi_0(R)\Phi_{nlm}(\vett R)
\ee
where $\Phi_0(R)\propto \exp(-\kappa R^2/2)$ and $\Phi_{nlm} (\vett
R)\propto R^l L_{n}^{l+1/2} (\kappa R^2) Y_{lm}(\theta,\phi)
\exp(-\kappa R^2/2)$ are the ground-state and
excited-state wave functions of a non-interacting gas trapped by an
harmonic (3D spherical) oscillator of frequency $\kappa$,
$L_{n}^{l+1/2}(\kappa R^2)$ being the
Laguerre polynomials. 
Consistent with this Ansatz, the Hartree term
vanishes for all higher multipoles, from orthogonality of
$\Phi_{nlm}(\vett R)$  to the
wave function $R_{\alpha} \Phi_0(R)$ of the dipolar excitation. We are
therefore left with 
the following equation of motion for the density fluctuation:
\bea
\omega^2 n_1(\vett R,\omega)&=&\frac{1}{4}\nabla_{\vett
R}\cdot \left\{ n_0(\vett R) \nabla_{\vett R} \left[\frac{1}{\sqrt{n_0(\vett 
R)}}\nabla^2 \frac{n_1(\vett R,\omega)}{\sqrt{n_0(\vett 
R)}}\right.\right.\nonumber \\&-& \left.\left.\frac{n_1(\vett
R,\omega)}{(n_0(\vett R))^{3/2}}\nabla^2 
\sqrt{n_0(\vett R)}\right]\right\}\;.
\eea
Since the equilibrium density is already of the form appropriate to
the ground state of an harmonic oscillator, Eq.~(20) is identically
satisfied by the Ansatz made in Eq.~(19). Therefore, the frequencies
of the modes with $n\ge 2$ are $\omega=n\kappa$, as anticipated. 

\section{Summary}

In summary, we have shown that for the T = 0 limit of the N-Boson
model in a parabolic potential and with harmonic interparticle
interactions, the linearized equation of motion involves kinetic
contributions of the kind exhibited in Eq.~(11). The corresponding
Gross-Pitaevskii equation (7) for the equilibrium density profile has
a solution which agrees with the weak-coupling limit of the exact
profile given in ref.~[5]. Inputting this equilibrium profile in
Eq.~(11) enables the dispersion relation of the collective excitations
to be obtained analyticaly. The frequency of the dipole mode is, in
accord with the Kohn theorem [7 - 9], that of the bare trap. The
frequencies of the higher-multipole modes are directly related to the
trap frequency renormalized by the interactions. 

\ack
One of us (NHM) wishes to acknowledge generous hospitality from the
Scuola Normale Superiore di Pisa during the period in which his
contribution to this study was completed.

\end{document}